\documentstyle[12pt]{article}
\input psfig
\def\be{\begin{equation}}
\def\ee{\end{equation}}
\def\ba{\begin{array}}
\def\ea{\end{array}}
\def\beqn{\begin{eqnarray}}
\def\eeqn{\end{eqnarray}}  
\def\bt{\begin{tabular}}
\def\et{\end{tabular}}
\def\bc{\begin{center}}
\def\ec{\end{center}}

\begin{document}
 \title{Three flavor neutrino oscillations, LSND, SNP and ANP}
 \author{Gulsheen Ahuja, Monika Randhawa and Manmohan Gupta \\
 {\it Department of Physics,} \\
 {\it Centre of Advanced Study in Physics,} \\
 {\it Panjab University,Chandigarh-160 014, India.}}   
 \maketitle
 \begin{abstract}
 The recently reanalysed LSND data is investigated in the context of
 three flavor oscillations, with an emphasis on mass hierarchies and
 $s_{13}$. The resultant mass hierarchies and oscillation angles are
 tested with regard to ``key" features of solar neutrinos and
 atmospheric neutrinos, e.g., ``average" survival probability in the
 case of solar neutrinos, and zenith angle dependence and up-down
 asymmetry in the case of high energy atmospheric neutrinos. We find
 there are three distinct mass hierarchies, e.g.,
 $\Delta m << \Delta M, \Delta m < \Delta M$ and
 $\Delta m \simeq \Delta M$. In the first and second case, the
 calculated range of $s_{13}$ is in agreement with the ``LMA" solution
 of Akhmedov {\it et al.}, the lower limit on $s_{13}$ in these cases
 is also in agreement with the recent analysis of Garcia {\it et al.}
 based on the constraints of SNP, ANP and CHOOZ, therefore, strongly
 supporting the neutrino oscillations observed at LSND. Further, the
 solutions of $s_{13}$ found in the third case correspond to the
 value of $s_{13}$ found by Akhmedov {\it et al.} in the case of
 ``SMA" and ``LOW" solutions.  A rough estimate of the possibility of
 the existence of CP violation in the leptonic sector is also carried
 out for different possible ranges of $s_{13}$, indicating that the CP
 asymmetries may be measureable even in the case of LSND.
 \end{abstract}
 Neutrino oscillations provide the preferred solution for the
 anomalous behaviour observed in the case of solar neutrinos
 \cite{sol-h}-\cite{sol-sn}, atmospheric neutrinos
 \cite{at-k}-\cite{at-ma}, as well as at LSND \cite{lsnd}. This
 essentially implies that the neutrinos are massive particles and the
 observed flavor eigenstates are linear combinations of mass
 eigenstates, in parallel to the quark mixing phenomenon. By far, the
 strongest evidence for neutrino oscillations is provided by the
 SuperKamiokande(SK) atmospheric neutrino data \cite{at-sk}. In view
 of the fact that the signal for neutrino oscillations is also the
 signal for physics beyond the Standard Model (SM)
 \cite{gla,wein,sal}, an intense amount of activity
 \cite{bil}-\cite{akh}, both at the experimental as well as at the
 phenomenological level, is going on to fix the mass hierarchies as
 well as the parameters of the neutrino oscillations.

    Several detailed and exhaustive analyses \cite{spec} have been
 carried out in the case of the solar neutrino data as well as in the
 case of the atmospheric neutrino data. These analyses along with the
 results from several other experiments have provided valuable
 information about the masses as well as about the mixing parameters.
 The constraints on masses and mixings are presented in terms of the
 mixing angles $( \theta_{12}, \theta_{13}, \theta_{23})$, defined
 analogous to the quark mixing angles \cite{pdg}, and the mass square
 differences. The most important constraint is provided by the SK
 analysis of their atmospheric neutrino data, \cite{at-sk,fog,gar},
 for example,
 \be  \Delta m_{32}^{2} \simeq(2-6) \times 10^{-3} eV^2,
       {\rm sin} \theta_{23} \simeq 0.545-0.839,    \label{atdat}  \ee
 where subscripts `2' and `3' correspond to $ \nu_{ \mu}$ and
 $ \nu_{ \tau}$. Although the constraint is for two flavor oscillations,
 however, in view of the CHOOZ constraint \cite{chooz}
 \be  {\rm sin}^{2}\theta_{13} \equiv |U_{e3}|^{2} \leq (0.06-0.018)~
 {\rm for}~ \Delta m_{31}^{2}=(1.5-5) \times 10^{-3} eV^2,  \label
   {cho}  \ee
 it remains valid in the case of three flavor oscillations also. It is
 generally believed that the bimaximal mixing matrix \cite{kay}
 provides a good approximation to the neutrino mixing matrix and is
 also consistent with the recent most exhaustive three flavor analysis of
 Solar Neutrino Problem (SNP) and Atmospheric Neutrino Problem (ANP)
 by Garcia {\it et al.} \cite{gar}.

    It is widely believed that the oscillation solutions to SNP, ANP
 and LSND data requires three different mass scales \cite{giu},
 therefore a simultaneous solution seems to be possible only within
 four flavor(3 active + 1 sterile) scenario \cite{giu}. However, the
 latest analysis of the SK data \cite{sol-sk} strongly indicates the
 absence of the role of sterile neutrinos in understanding SNP and ANP.
 This has complicated the issues related to neutrino mass scales.
 Besides this, there are several issues where more information is needed
 to deepen the understanding of the phenomenology of neutrino
 oscillations. One of the important issue is to find the lower limit of
 $s_{13}$, the upper limit, of course, is provided by the CHOOZ
 constraint. Some attempts have been made in this direction
 \cite{gar,abr}, however a clear picture is yet to emerge. Similarly,
 despite considerable progress in understanding the neutrino mass
 differences, more information is very much desirable in knowing the
 various mass hierarchies which can fit the solar neutrino data, the
 atmospheric neutrino data and the LSND data simultaneously. The other
 burning issue, in the context of neutrino oscillations, is the
 existence or non existence of CP violation in the leptonic sector,
 for which serious effort is going on to decipher it in the Long
 BaseLine(LBL) experiments. Any clue about CP violation, from
 phenomenological considerations, would have far reaching consequences
 for the neutrino oscillation phenomena as well as for the planned LBL
 experiments and Neutrino Factories \cite{bar,ell,akh,mina}.

 The reaffirmation of the LSND results presented at Neutrino 2000 as
 well as the lowering of their probability, would have important
 consequences as this is an appearance experiment, unlike solar
 neutrino experiments and atmospheric neutrino experiments. Another
 extremely important observation which has been emphasized at Neutrino
 2000 is the absence of energy dependence of solar neutrino recoil
 energy spectrum by SK \cite{sol-sk} as well as by Sudbery Neutrino
 Observatory (SNO) \cite{sol-sn}. Similarly, the preliminary
 observations of K2K experiment \cite{k2k} as well as of SNO
 endorse the conclusions of
 SK regarding the hypothesis of neutrino oscillations. In view of the
 recent observations of SK about sterile neutrinos as well as the
 issue of three different mass scales being required to explain SNP,
 ANP and LSND results, the question regarding the number of neutrino
 species has become extremely important. In this context, Minakata
 \cite{mina} has strongly advocated that one should try to re-explore,
 within three active flavors, the crucial features of the solar neutrino
 data, the atmospheric neutrino data and the LSND data which can be
 reconciled simultaneously.
 Some preliminary attempts \cite{gab,con,tesh} have been made in
 this connection which need to be updated in the context of the
 observations at Neutrino 2000. Further, these attempts have not gone
 into the details of the mixing angles, in particular the lower limit
 of $s_{13}$ and its implications on CP violation. Furthermore, these
 analyses do not adequately emphasize the issues related to flux
 uncertainties as well as up-down asymmetry in the case of atmospheric
 neutrinos.

    The purpose of the present communication, on the one hand, is to
 explore, within three active flavors, the possibility of reconciling
 the latest LSND data with some of the ``key" features of solar and
 atmospheric neutrino data. On the other hand, it is to explore
 the constraints on the element $s_{13}$, the resultant mass
 hierarchies as well as the possible clues about the order of CP
 violation in the leptonic sector.

    Before we proceed further, we underline the ``key" features of the
 solar neutrinos and the atmospheric neutrinos which are relevant to
 the present investigation. In the case of solar neutrinos, the most
 important feature of the data is the ratio of flux of solar neutrinos
 observed by various ongoing experiments to the flux predicted  by
 Standard Solar Model(SSM) \cite{bah}. In view of the observation at
 SK and SNO about the energy independence of the solar neutrino flux,
 it is expected that the above mentioned ratio should be same in all
 the solar neutrino experiments. In this situation, as advocated by
 Bahcall {\it et al.} \cite{bah}, and Conforto {\it et al.} \cite{con},
 one can consider the ``average" solar survival probability as,
   \be     P_{sol}=0.50 \pm 0.06.   \label{cpsol}    \ee

    Similarly, in the case of atmospheric neutrinos the ``key" aspects
 of the data which indicate the possibility of neutrino oscillations
 are zenith angle dependence of the ratio (observed versus Monte Carlo
 (MC)) of the neutrino fluxes and the up-down asymmetry. In view of
 the fact that the ratios are very much dependent on the MC
 simulations of neutrino fluxes, it is perhaps desirable to
 concentrate on those features of the data wherein the flux
 uncertainties are minimum. Details in this regard will be discussed
 later in the text.

    To begin with, we consider the neutrino mixing matrix, 
  \be \left( \ba {c} \nu_e \\ \nu_{\mu} \\ \nu_{\tau} \ea \right)
  = \left( \ba{ccc} U_{11} & U_{12} & U_{13} \\ U_{21} & U_{22} &
  U_{23} \\ U_{31} & U_{32} & U_{33} \ea \right) 
 \left( \ba {c} \nu_1\\ \nu_2 \\ \nu_3 \ea \right),  \label{nm}  \ee
 where $ \nu_{e}$, $ \nu_{\mu}$, $ \nu_{\tau}$ are the flavor
 eigenstates and $ \nu_1$, $ \nu_2$, $ \nu_3$ are the mass
 eigenstates. In the PDG representation \cite{pdg}, the mixing matrix
 can be expressed as,
  \be U=\left( \ba{ccl} c_{12} c_{13} & s_{12} c_{13} &
  s_{13}e^{-i \delta_{13}} \\ - s_{12} c_{23} - c_{12} s_{23}
  s_{13} e^{i \delta_{13}} & c_{12} c_{23} - s_{12} s_{23}
  s_{13} e^{i \delta_{13}} & s_{23} c_{13} \\ s_{12} s_{23} - c_{12}
  c_{23} s_{13} e^{i \delta_{13}} & - c_{12} s_{23} - s_{12} c_{23}
  s_{13} e^{i \delta_{13}} & c_{23} c_{13} \ea \right). \label{mm} \ee

    In the sequel, we detail the relevant appearance and disappearance
 probabilities in terms of masses and mixing  angles pertaining to the
 LSND experiment, the SNP and the ANP. In the case of LSND experiment,
 we express the probability in terms of the mixing matrix. The
 expression for $P_{LSND}$, ignoring CP violation effects, can be
 expressed as \cite{plsnd}
 \beqn P_{LSND}&=& -4 U_{21} U_{11} U_{22} U_{12}
            {\rm sin}^2 \left( \frac{\Delta m^2 L} {4E}\right) \nonumber \\   
              & &   -4 U_{21} U_{11} U_{23} U_{13}
           {\rm sin}^2 \left( \frac{( \Delta m^2 + \Delta M^2) L} {4E}
                    \right) \nonumber \\
            & &     -4 U_{22} U_{12} U_{23} U_{13}
        {\rm sin}^2 \left( \frac{\Delta M^2 L} {4E}\right), \label{pl} \eeqn
 where $\Delta m^{2} \equiv \Delta m_{21}^{2}= m_{2}^{2}-m_{1}^{2}$,
 $\Delta M^{2} \equiv \Delta m_{32}^{2}= m_{3}^{2}-m_{2}^{2}$, $L$
 denotes the neutrino flight path $i.e.$ the distance between the
 neutrino source and the dectector and $E$ is the average energy of
 the neutrinos. If $\Delta m^2, \Delta M^2$ are expressed in $eV^2$
 and $L$ is expressed in metres then $E$ is in $MeV$. The LSND set up
 is characterised by $L=30$m and $36 MeV < E < 60 MeV$. For the
 purpose of calculations, we take the average energy $E=42 MeV$ and
 from their latest analysis \cite{lsnd}, we have
  \be     P_{LSND}=(2.5 \pm 0.6 \pm 0.4) \times 10^{-3}~
{\rm with}~ \Delta m^2 >0.2 eV^2.  \label{pln}      \ee
 It is to be noted that when the above constraint is combined with the
 limit provided by the CDHSW data \cite{cdhsw}, the range of
 $\Delta m_{LSND}^2$ becomes
 \be   0.2 eV^2 < \Delta m_{LSND}^2 < 0.4 eV^2.    \label{lsmr}  \ee

    In view of the fact that $s_{12}$ and $s_{23}$ are fairly well
 known, one can make use of the equation~(\ref{pl}), with the
 identification of $ \Delta M^{2}$ with $\Delta m_{LSND}^2$, to find
 the element $s_{13}$ for different mass hierarchies, however these
 values have to satisfy the ``key" features of the solar neutrinos
 as well as the atmospheric neutrinos.

    To this end, we first consider the case of solar neutrinos. The
 survival probability of solar electron neutrinos is given by,
 \beqn P_{sol}&=& 1-4 U_{11} U_{11} U_{12} U_{12}
             {\rm sin}^2 \left( \frac{\Delta m^2 L}{4E}\right)  \nonumber \\
                  & &  -4 U_{11} U_{11} U_{13} U_{13}
   {\rm sin}^2 \left( \frac{( \Delta m^2 + \Delta M^2) L}{4E}
                   \right) \nonumber \\
           & &   -4 U_{12} U_{12} U_{13} U_{13}
        {\rm sin}^2 \left( \frac{\Delta M^2 L}{4E}\right).  \label{ps} \eeqn
 The extremely large $L/E$ factor, as well as the range of
 $\Delta m^{2}$ considered here, enables one to simplify the above
 expression by carrying out averaging, which in terms of the mixing
 angles becomes
 \be P_{sol}=1- 2 c_{13}^2 ( s_{13}^2 + s_{12}^2 c_{12}^2 c_{13}^2).
               \label{aps}   \ee  

    In the case of atmospheric neutrinos, the most important feature
 is the zenith angle dependence of the ratio of the observed neutrinos
 to the MC expectations. For example, the ratios in the case of
 electron neutrinos ($R_{e}$) and muon neutrinos ($R_{ \mu}$) are
 defined as follows,
     \be   R_e=P_{ee}+r P_{ \mu e},   \label{ref}     \ee  
 \be  R_ \mu =P_ {\mu \mu} + \frac{1}{r} P_{ \mu e},  \label{ruf}  \ee
 where $P_{ee}$, $P_{ \mu \mu}$ are the survival probabilities,
 $P_{ \mu e}$ is the transition probability, and
       \be        r=  \frac{N_{ \mu}^o}{N_{e}^o}     \label{r}   \ee
 is the expected ratio of fluxes without oscillations (where
 $N_{ \mu (e)}^o$ is the initial flux of muon (electron) neutrinos).
 The ratio $r$ varies from 1.6 ( for sub GeV neutrinos ) to 3.0 ( for
 multi GeV neutrinos ). Further, this ratio depends on the zenith
 angle also, varying from 1.6 for the horizontal neutrinos to 3.0 for
 the vertical ones. It is obvious that the uncertainties in $r$ are
 bound to affect the calculation of the zenith angle dependence of
 $R_{e}$ and $R_{ \mu}$. Since the purpose of the present
 communication is not to  go into the extensive details of these
 aspects of data, therefore we would like to concentrate on that
 part of the data which is free from such uncertainties. Interestingly,
 it is to be noted that these uncertainties get highly suppressed for
 the case of high energy upward going neutrinos. It is also to be
 noted that the SK detector depicts better correlation for upgoing
 high energy neutrinos with the charged leptons produced in the
 detector. Further, the asymmetry $A$, defined as
        \be   A=\frac{U-D}{U+D},  \label{a}     \ee 
 where $U$ is the number of upward going events with zenith angles
 in the range $ -1<{\rm cos} \theta<-0.2 $ and $D$ is the number of
 downward going events with $ 0.2< {\rm cos} \theta<1 $, is better
 defined for high energy neutrinos \cite{akh}.

    In view of the above mentioned reasons, it is desirable to
 consider the zenith angle dependence of $R_{e}$ and $R_{ \mu}$ for
 high energy upgoing neutrinos. The zenith angle dependence is
 generally expressed in terms of $L/E$. It is to be noted that for the
 vertically upgoing neutrinos, the length $(L)$ traversed is $ \sim$
 13000 km. For the SK geometry, the horizontal neutrinos correspond to
 those travelling $ \sim$ 500 km, therefore, for the purpose of
 studying zenith angle dependence which does not involve uncertainty
 due to flux ratio $r$, it is desirable to avoid lengths
 $ \simeq 1000$ km.

    Keeping in mind the mentioned constraints, the three flavor
 probability expressions $P_{ee},~ P_{ \mu \mu }$ and $P_{ \mu e}$
 required to evaluate  $R_{e}$ and $R_{ \mu}$ can be expressed as
 follows. The electron neutrino survival probability is given by,
  \beqn P_{e e}&=& 1-4 U_{11} U_{11} U_{12} U_{12}
          {\rm sin}^2 \left( \frac{\Delta m^2 L}{4E}\right)  \nonumber  \\
         & &     -4 U_{11} U_{11} U_{13} U_{13}
         {\rm sin}^2 \left( \frac{( \Delta m^2 + \Delta M^2) L}{4E}
                \right)   \nonumber  \\
         & &     -4 U_{12} U_{12} U_{13} U_{13}
       {\rm sin}^2 \left( \frac{\Delta M^2 L}{4E}\right). \label{paes} \eeqn
 Similarly the muon neutrino survival probability is given by, \pagebreak
 \beqn P_{ \mu \mu }&=&1-4 U_{21} U_{21} U_{22} U_{22}
          {\rm sin}^2 \left( \frac{\Delta m^2 L}{4E}\right)   \nonumber \\
         & &     -4 U_{21} U_{21} U_{23} U_{23}
          {\rm sin}^2 \left( \frac{( \Delta m^2 + \Delta M^2) L}{4E}
                \right)   \nonumber \\
         & &     -4 U_{22} U_{22} U_{23} U_{23}
      {\rm sin}^2 \left( \frac{\Delta M^2 L}{4E}\right).  \label{paus} \eeqn
 Likewise, the transition probability is expressed as, 
 \beqn P_{ \mu e}&=& -4 U_{21} U_{11} U_{22} U_{12}
         {\rm sin}^2 \left( \frac{\Delta m^2 L}{4E}\right)  \nonumber  \\
            & &  -4 U_{21} U_{11} U_{23} U_{13}
         {\rm sin}^2 \left( \frac{( \Delta m^2 + \Delta M^2) L}{4E}
                \right)  \nonumber \\
           & &   -4 U_{22} U_{12} U_{23} U_{13}
       {\rm sin}^2 \left( \frac{\Delta M^2 L}{4E}\right). \label{pa}  \eeqn
 The factor $L/E$ is quite large for the upgoing neutrinos considered
 here, therefore, for the purpose of calculations the above
 expressions get reduced to the following,
  \be P_{ e e}=1-2 U_{11}^2 U_{12}^2-2 U_{11}^2 U_{13}^2  \
                            -2 U_{12}^2 U_{13}^2, \label{apaes}  \ee

  \be P_{\mu \mu}=1-2 U_{21}^2 U_{22}^2 -2 U_{21}^2 U_{23}^2
                            -2 U_{22}^2 U_{23}^2, \label{apaus}   \ee

  \be P_{ \mu e}= 2 U_{23}^2 U_{13}^2 - 2 U_{22} U_{12} U_{21} U_{11}.
                  \label{apa}      \ee

    In Tables \ref{tabnh} and \ref{tabde}, we have presented the results
 of our calculations, for different mass hierarchies, incorporating the
 constraints due to the LSND results as well as due to the ``key"
 features of SNP and ANP. To begin with, using the LSND probability
 given by the equation (\ref{pl}), we have calculated $s_{13}$. The
 values of $ \Delta M^2$ are constrained by the equation (\ref{lsmr}),
 the values of $s_{23}$ by the equation (\ref{atdat}) whereas the
 values of $s_{12}$ used for calculations are fairly consistent with
 bimaximal mixing as well as with the analysis of Garcia {\it et al.}
 \cite{gar}.

    In Table \ref{tabnh}, we have presented a representative sample of
 the calculations carried out in the case of ``natural" mass hierarchy
 $ \Delta m^2 \ll \Delta M^2$, with $\Delta m^{2}=10^{-4}eV^{2}$. The
 results corresponding to the hierarchy $ \Delta m^2 \ < \Delta M^2$
 are very much similar to those of $ \Delta m^2 \ll \Delta M^2$, hence
 are not presented in the tables. It is interesting to mention that
 while carrying out the analysis of the range of $s_{13}$, using the
 LSND constraint, we find that $s_{12}$ plays no role. However,
 because of the fact that $P_{sol}$ depends on $s_{12}$ values, it is
 not possible to consider values of $s_{12} < 0.6$. The fact that
 $P_{LSND}$ is independent of $s_{12}$ values can be easily understood
 if one carefully examines the equation (\ref{pl}). Since
 $\Delta m^2 \ll \Delta M^2$, the first term of the equation (\ref{pl})
 can be neglected and the contribution of $\Delta m^2$ from the second
 term can be ignored. The second and the third term can be combined,
 yielding in terms of mixing angles,
   \be P_{LSND}=4 s_{23}^2 c_{13}^2 s_{13}^2 {\rm sin}^2 \left( \frac
  { \Delta M^2 L}{4E} \right)~,   \label{splsnd}       \ee 
 which has no $s_{12}$ dependence. Therefore, while carrying out the
 calculations, we have considered the bimaximal mixing value of
 $s_{12}$. From Table \ref{tabnh}, it is quite evident that when we
 scan the possible range of $P_{LSND}$, $ \Delta M^2$ and $s_{23}$, we
 get the range of $s_{13}$ as $(0.07-0.29)$. From the table it should
 be noted that $P_{sol}$ is very much in agreement with equation
 (\ref{cpsol}) for values of $s_{13}$ only upto 0.18. Similarly, for
 the case of atmospheric neutrinos, $R_{e}$ and $R_{ \mu}$ also agree
 with the data \cite{bil} for the same upper limit of $s_{13}$.
 Therefore, when we consider the constraints imposed by SNP, ANP and
 the LSND experiment simultaneously, the range of $s_{13}$ gets
 restricted to $(0.07-0.18)$. It needs to be added that for the region
 of $L/E$ considered here, $R_{e}$ and $R_{ \mu}$ assume constant
 values because of averaging of various oscillatory terms in the
 probability expressions. 

    In Table \ref{tabde}, we have presented the results of the
 calculations corresponding to the ``degenerate" case, $i.e.$ where
 $\Delta m^2$ and $\Delta M^2$ are of the same order of magnitude. As
 a first step, we have tried to find the largest possible value of
 $ \Delta m^{2}$ which can fit the various constraints imposed by LSND,
 SNP and ANP. In this context, we find that a value of
 $ \Delta m^2 >$ 0.085 $eV^2$ is not able to fit $P_{LSND}$. It is
 also to be noted that the range of $s_{23}$ is very much limited to
 values $ \simeq$ 0.82, which is almost near the largest value
 admitted by SK. Using these values of $ \Delta m^2$ and $s_{23}$, we
 have calculated possible values of $s_{13}$ as well as of $s_{12}$
 which fit the data. As is evident from the table, the values of
 $s_{12}$ are also quite limited to the range $0.66-0.70$, whereas,
 $s_{13}$ takes on much smaller values, $0.001-0.03$, as compared to
 the previous case where $\Delta m^2 \ll \Delta M^2$.

    A closer scrutiny of our results reveals several points. It needs
 to be emphasized that the lower value of $s_{13}$ found in the case
 $\Delta m^2 \ll \Delta M^2$, is in agreement with the value found
 recently by Garcia {\it et al.} \cite{gar}, by carrying out a
 simultaneous analysis of solar, atmospheric and CHOOZ data.
 Interestingly, the range of $s_{13}$ found by our calculations is in
 complete agreement with the range found by Akhmedov {\it et al.}
 \cite {abr}, in the case of ``LMA" solution of the SNP. Further, the
 upper range of $s_{13}$, surprisingly, is in agreement with that of
 CHOOZ constraint, however, the present upper limit, unlike the CHOOZ
 case, is scale independent. Similarly, in the second case where
 $\Delta m^2 \simeq \Delta M^2$, the calculated values of $s_{13}$
 correspond to the value of $s_{13}$ found by Akhmedov {\it et al.}
 \cite{abr}, in the case of ``SMA" and ``LOW" solutions. This
 agreement of $s_{13}$ range found primarily from LSND data with other
 diverse analyses lends strong support to the neutrino oscillations
 observed at LSND.

    In the above mentioned tables, we have presented the results for
 two different mass hierarchies, e.g., $\Delta m^2 \ll \Delta M^2$ and
 $\Delta m^2 \simeq \Delta M^2$. In the latter case, we have already
 mentioned a specific value of $\Delta m^2$ which in fact is the
 highest value fitting the LSND data. In the first case, we have given
 a typical value, however our results are very much valid if
 $\Delta m^2$ corresponds to either the ``LMA" or the ``LOW" solutions
 to the SNP. The calculations also remain valid even if
 $\Delta m^2 \simeq 10^{-2}$ $eV^2$.

    In the tables, we have not included the results of the asymmetries
 calculated from equation~(\ref{a}). In case we impose the range of
 $A$ within $1 \sigma$, e.g.,- 0.35 to - 0.25 \cite{at-sk}, we find that
 the results corresponding to the degenerate case are within this
 range, however in the case  $\Delta m^2 \ll  \Delta M^2$ only values
 of $s_{13} \leq$ 0.1 satisfy the above mentioned range of the
 asymmetry.

    It is interesting to mention that the present calculations for
 both the mass hierarchies are fully compatible with the $ \nu_{ \mu}$
 disappearance observations at K2K. Although the K2K results show up
 only at $2 \sigma$ level, still their compatibility with the present
 calculations supports the oscillation hypothesis.

    In the present calculations we have not considered the downgoing
 atmospheric neutrinos, primarily because of the flux uncertainties
 associated with these neutrinos as well as because of the fact that
 high energy neutrinos do not oscillate for these lengths. However, it
 may be mentioned that for the low energy neutrinos $ \Delta m^{2} L/E$
 can be $\sim 1$, in which case, the calculations become highly
 scale sensitive.

    It is interesting to explore which additional features as well as
 mass hierarchies, apart from the ``key" features of SNP and ANP,
 would be describable if one extends the present 3 active flavors to
 3 active + 1 sterile scenario. The details of these will be published
 elsewhere.

\begin{figure}
   \centerline{\psfig{figure=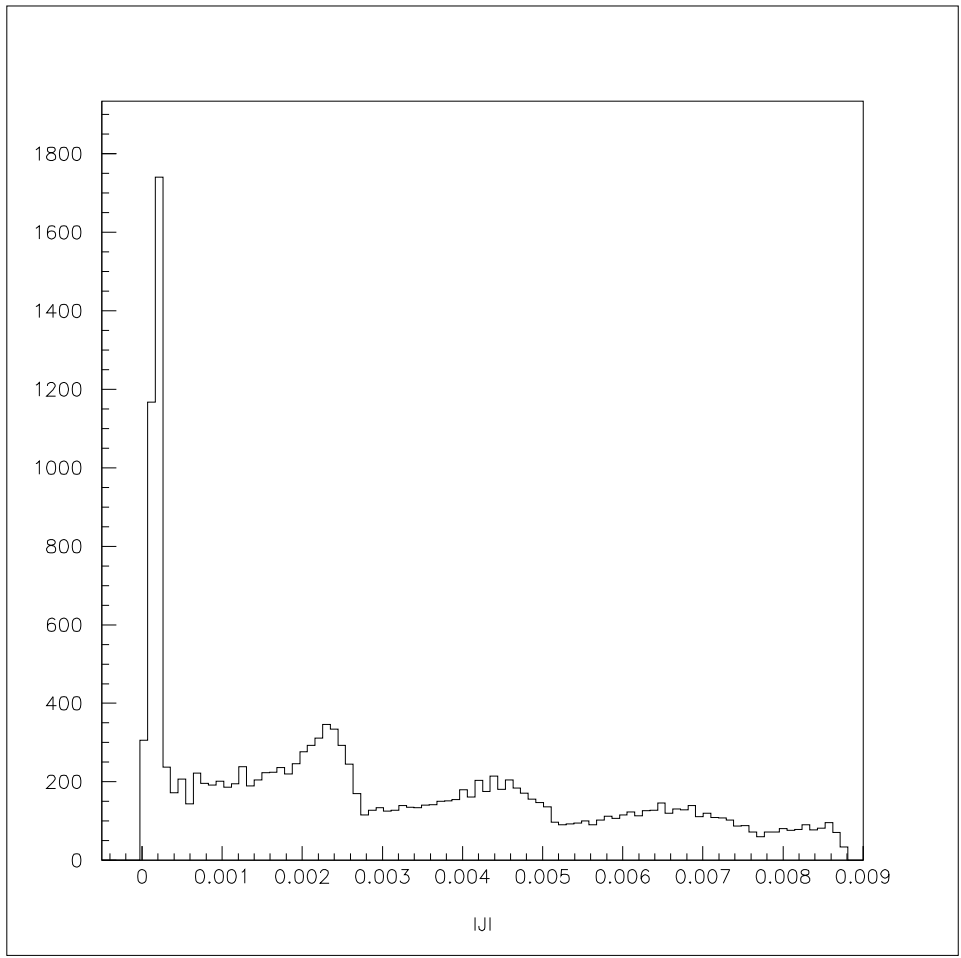,width=5in,height=5in}}
   \caption{Histogram of $|J|$ plotted
   by considering the orthogonality relationship of I and II row of
   the mixing matrix.}
  \label{figj}
  \end{figure}

    After having found mixing angles which are compatible with the
 three neutrino experiments, it is desirable to examine the kind of
 texture specific mass matrices which would reproduce these mixing
 angles, as well as the possibility of CP violation in the leptonic
 sector. Regarding the mass matrices,
 some preliminary investigations
 have been carried out earlier \cite{masmat}, details in the present
 context would be published elsewhere. Regarding CP violation,
 assuming that $\delta$ is nonzero in the leptonic mixing matrix, one
 can estimate the order of CP violation, by calculating Jarlskog's
 rephasing invariant parameter J \cite{jarl}, through the relationship
 \be |J|=2 \times {\rm~ area~ of~ the~ unitarity~ triangle},
             \label{ar} \ee
 for details we refer the reader to reference \cite{mr1}. In analogy
 to the quark sector, $J$ can be evaluated by using any of the six
 unitarity triangles in the leptonic sector, through
 equation (\ref{ar}). A nonzero area of any of the unitarity triangle
 would imply CP violation, which we ensure by putting the constraint
 that the sum of the two sides of the triangle is always greater than
 the third side. Using this constraint $J$ can be evaluated for any
 of the six unitarity triangles, however we will present the results
 corresponding to the unitarity triangle formed by considering the
 orthogonality of the I and II row of the mixing matrix.  In figure
 \ref{figj}, we have plotted the probability distribution of $J$ by
 varying $s_{12},~ s_{23}$ and  $s_{13}$ within the ranges mentioned
 in the I row of the Table \ref{tabcp} and $\delta$ in the range
 0$^{\rm o}$ to 180$^{\rm o}$. Interestingly the histogram shows a
 sharp peak around $J=2.126 \times 10^{-4}$, a corresponding guassian
 fit results in the range
    \be J=(2.126 \pm 0.272)\times 10^{-4}. \label{j1} \ee
 The corresponding  most probable range for $\delta$ is
  \beqn \delta&=& 1^{\rm o}~ {\rm to}~ 10^{\rm o}~~~~~~~~~~~~~
 ({\rm in~ I~ quadrant}),   \nonumber \\
 & & 170^{\rm o}~ {\rm to}~ 180^{\rm o}~~~~~~~~~({\rm in~ II~
                                 quadrant}).  \label{d1} \eeqn
 We have repeated the above procedure for the  mixing angles listed in
 row II and III of Table \ref{tabcp}. The corresponding $J$ and
 $\delta$ are also listed in the table. In the case
 $s_{13}= 0.001-0.10$, we have mentioned for $ \delta$ the value
 $90^{\rm o} \pm 14^{\rm o}$, however, $ \delta$ can also exist with
 significant probability in the range $1^{\rm o}~ {\rm to}~ 5^{\rm o}$
 in the first quadrant and correspondingly
 $175^{\rm o}~ {\rm to}~ 180^{\rm o}$ in the second quadrant.

    It is interesting to consider the CP asymmetry in the case 
 $\Delta M^2 \simeq \Delta m^2$.  The CP asymmetry \cite{ell} is
 defined as follows
 \be A_{CP}~=~ \frac{P( \nu_{ \mu} \rightarrow \nu_{e})~-~
         P( \overline{ \nu}_{ \mu} \rightarrow \overline{ \nu}_{e})}
         {P( \nu_{ \mu} \rightarrow \nu_{e})~+~
         P( \overline{ \nu}_{ \mu} \rightarrow \overline{ \nu}_{e})} 
         \simeq~ \frac{4{\rm sin}^2 \theta_{12}{\rm sin} \delta}{{\rm sin} \theta_{13}}
    {\rm sin} \left( \frac{2 \Delta m^2_{12} L}{4E} \right). \label{a1}  \ee
 For the degenerate case, as well as using the LSND parameters $L/E$,
 we get a rough estimate of $A_{CP}$ as follows,
 \be A_{CP} \simeq {\rm sin}~\delta  \label{acp}. \ee
 As shown in the Table \ref{tabcp}, for this case $\delta$ can take
 values $ 1^{\rm o}~ {\rm to}~ 10^{\rm o}$, therefore the asymmetry
 can be reasonable even  for short baseline experiment like LSND,
 however in the case of LBL experiments the asymmetry can take large
 values. Similarly, in the case of natural mass hierarchy, the largest
 value of $A_{CP}$ for LBL experiments, for example K2K experiments,
 can be given as
 \be  A_{CP} \simeq 0.4 . \label{acp1} \ee

    In conclusion, we would like to emphasize that we have carried out
 a simultaneous analysis, with three flavor oscillations, of the
 recently updated LSND data along with the ``key" features of solar
 neutrinos and atmospheric neutrinos. Interestingly, we find that
 there are three mass hierarchies which are able to fit the LSND data,
 the ``averaged" solar neutrino probability and the zenith angle
 dependence for high energy neutrinos. The mass hierarchies found are
 $\Delta m^2 \ll \Delta M^2$, $\Delta m^2 < \Delta M^2$ and
 $\Delta m^2 \simeq \Delta M^2$, the latter being valid only for
 values of $s_{23} \simeq 0.82$. Out of these three mass hierarchies,
 the first and the second possibilities give similar results. In a
 particular case with $\Delta m^2=10^{-4}eV^2$, we find the range of
 $s_{13}$ to be $0.07-0.18$. This range is in fair agreement with the
 ``LMA" solution of Akhmedov {\it et al.} \cite{abr}. Further, the
 lower limit of this range is in agreement with the most recent and
 exhaustive analysis by Garcia {\it et al.} \cite{gar} based on a
 simultaneous analysis of solar neutrino data, atmospheric neutrino
 data and CHOOZ results. Furthermore, the range of $s_{13}$, for
 example $0.001-0.032$, found in the degenerate case corresponds to
 the values of $s_{13}$ found by Akhmedov {\it et al.} \cite{abr} in
 the case of ``SMA" and ``LOW" solutions. These observations, in a way,
 show the compatibility of the oscillations observed at LSND with the
 solar and the atmospheric neutrino data. If in the case of
 atmospheric data, the up-down asymmetry is also included in our
 analysis, then the solution space gets restricted if the results of
 asymmetry are included upto $1 \sigma$ level. However, if the results
 are included upto $2 \sigma$ level, then most of the solutions remain
 valid. We have also made a rough estimate of CP violating asymmetries
 assuming the existence of CP violation in the leptonic sector.
 Interestingly, CP violation may be observable even at LSND.
 
 \vskip 0.2cm {\bf Acknowledgements} \\
 GA and MG would like to thank S.D. Sharma for useful discussions.
 MR would like to thank CSIR, Govt. of India for financial support.
 GA and MR would like to thank the Chairman, Department of Physics
 for providing facilities to work in the department.
  \pagebreak

\begin{table}
\begin{tabular}{|c|l|l|l|l|l|l|l|l|}  \hline
$P_{LSND} (\times 10^{-3})$ & $\Delta m^2$ & $\Delta M^2$ & $s_{23}$
& $s_{13}$ & $P_{atm}$ & $P_{sol}$ & $R_{e}$ & $R_{ \mu}$ \\ \hline
1.8 & $10^{-4}$ & 0.40 & 0.82 & 0.07 & 0.33 & 0.50  & 1.00 & 0.56 \\
2.0 & $10^{-4}$ & 0.38 & 0.78 & 0.08 & 0.35 & 0.49  & 1.01 & 0.52  \\
2.2& $10^{-4}$  & 0.35 & 0.74 & 0.10 & 0.36 & 0.49  & 1.01 & 0.51 \\
2.4 & $10^{-4}$ & 0.32 & 0.70 & 0.12 & 0.36 & 0.49  & 1.01 & 0.50 \\
2.6 & $10^{-4}$ & 0.29 & 0.66 & 0.15 & 0.35 & 0.48  & 1.01 & 0.52 \\
2.8 & $10^{-4}$ & 0.26 & 0.62 & 0.18 & 0.33 & 0.47  & 1.01 & 0.54 \\
3.0 & $10^{-4}$ & 0.23 & 0.58 & 0.23 & 0.30 & 0.45  & 1.00 & 0.58  \\
3.2 & $10^{-4}$ & 0.20 & 0.54 & 0.29 & 0.27 & 0.42  & 0.98 & 0.62 \\ \hline
\end{tabular}
\caption{Calculated values of $s_{13},~P_{atm},~P_{sol},~R_{e},
~R_{ \mu}$ in terms of $s_{12},~ s_{23}, ~\Delta m^2 (eV^2),
~\Delta M^2(eV^2)$, with $s_{12}=0.70$.}
\label{tabnh}
\end{table}

\begin{table}
\begin{tabular}{|c|l|l|l|l|l|l|l|l|l|}  \hline
$P_{LSND} (\times 10^{-3})$ & $\Delta m^2$ & $\Delta M^2$ &
$s_{12}$ & $s_{23}$ & $s_{13}$ & $P_{atm}$ & $P_{sol}$ & $R_{e}$
& $R_{ \mu}$ \\ \hline
2.1 & $0.085$ & 0.40 & 0.70 & 0.82 & 0.001 & 0.33 & 0.50  & 0.99 & 0.56  \\
2.4 & $0.085$ & 0.35 & 0.69 & 0.82 & 0.007 & 0.33 & 0.50  & 0.99 & 0.56 \\
2.7 & $0.085$ & 0.30 & 0.68 & 0.82 & 0.014 & 0.33 & 0.50  & 0.99 & 0.56  \\
3.0 & $0.085$ & 0.25 & 0.67 & 0.82 & 0.023 & 0.33 & 0.50  & 0.99 & 0.56  \\
3.2 & $0.085$ & 0.20 & 0.66 & 0.82 & 0.032 & 0.33 & 0.51  & 0.99 & 0.56  \\ \hline
\end{tabular}
\caption{Calculated values of $s_{13},~ P_{atm},~ P_{sol},~ R_{e},
~R_{ \mu}$ in terms of $s_{12},~ s_{23}, ~\Delta m^2 (eV^2),
~\Delta M^2(eV^2)$.}
\label{tabde}
\end{table}

\begin{table}
\begin{tabular}{|c|c|c|c|c|}  \hline
$s_{12}$ & $s_{23}$ & $s_{13}$ & $J$ & $\delta$ \\ \hline
&&&&\\
0.5 - 0.7 & 0.54 - 0.82 & 0.001 - 0.05 & $(2.104 \pm 0.262) \times 10^{-4}$
& $\left( \ba{c}   1^{\rm o}~ {\rm to}~ 10^{\rm o},\\
 170^{\rm o}~ {\rm to}~ 180^{\rm o} \ea \right)$ \\
&&&&\\
0.5 - 0.7 & 0.54 - 0.82 & 0.001 - 0.10 & $(2.382 \pm 0.079) \times 10^{-4}$
& $90^{\rm o} \pm 14^{\rm o}$ \\
&&&&\\
0.5 - 0.7 & 0.54 - 0.82 & 0.05 - 0.15 & $(1.179 \pm 0.650) \times 10^{-2}$
& $ 1^{\rm o}~ {\rm to}~ 180^{\rm o}$ \\ 
&&&&\\ \hline
\end{tabular}
\caption{$J$ and $\delta$ corresponding to different ranges of $s_{13}$.}
\label{tabcp}
\end{table}

\end{document}